\documentclass[aps, prc, reprint, superscriptaddress, nofootinbib, floatfix]{revtex4-2}
\usepackage[T1]{fontenc}
\usepackage[utf8]{inputenc}
\usepackage{graphicx}
\usepackage{amsmath,amssymb,bm}
\usepackage{dcolumn}
\usepackage{xcolor}

\definecolor{custompurple}{HTML}{9300d3}
\definecolor{customgreen}{HTML}{019d73}
\definecolor{customblue}{HTML}{57b5e8}
\usepackage[colorlinks=true,linkcolor=customblue, citecolor=blue,urlcolor=blue]{hyperref}

\begin{document}

\title{Responses of multiparticle observables to multidimensional nuclear deformation in relativistic heavy-ion collisions}

\author{Ying Shan Zhao}
\affiliation{State Key Laboratory of Dark Matter Physics, Shanghai Key Laboratory for Particle Physics and Cosmology,
Key Laboratory for Particle Astrophysics and Cosmology (MOE),
School of Physics and Astronomy, Shanghai Jiao Tong University, Shanghai 200240, China}

\author{Yifeng Sun}
\email{Contact author: sunyfphy@sjtu.edu.cn}
\affiliation{State Key Laboratory of Dark Matter Physics, Shanghai Key Laboratory for Particle Physics and Cosmology,
Key Laboratory for Particle Astrophysics and Cosmology (MOE),
School of Physics and Astronomy, Shanghai Jiao Tong University, Shanghai 200240, China}


\begin{abstract}
Relativistic heavy-ion collisions provide a unique opportunity to probe ground-state nuclear structure through its imprint on the initial collision geometry. We investigate how multiparticle observables respond to combined variations of quadrupole deformation, triaxiality, and hexadecapole deformation, using $^{129}$Xe+$^{129}$Xe collisions as a representative testing ground. We perform a joint analysis in the three-dimensional $(\beta_2,\gamma,\beta_4)$ parameter space and construct initial-state estimators for several flow and mean transverse momentum correlation observables. At the initial-state level, $\rho_2$ is primarily sensitive to $\beta_2$ and $\gamma$, with its sensitivity to $\gamma$ enhanced at nonzero $\beta_2$. The nonlinear response coefficient $\chi_{4,22}$ is predominantly sensitive to $\beta_4$, while its dependence on $\beta_2$ and $\gamma$ remains comparatively weak. Higher-order correlators exhibit more complex multidimensional response patterns; in particular, $\rho_{224}$ shows a dependence on $\gamma$ and $\beta_4$ that becomes more pronounced at finite $\beta_2$. We further employ the iEBE-VISHNU hybrid model to examine whether these deformation sensitivities survive the subsequent dynamical evolution. The final-state calculations indicate that the sensitivity of $\rho_2$ to $\beta_2$ and $\gamma$ is largely preserved, whereas the $\beta_4$ sensitivity of $\chi_{4,22}$ is substantially reduced. For the other higher-order observables, the initial-state sensitivities are modified by the evolution or cannot be resolved with the present statistics.
\end{abstract}

\maketitle

\section{Introduction}

Relativistic heavy-ion collision experiments at the Relativistic Heavy Ion Collider (RHIC) and the Large Hadron Collider (LHC) provide a unique opportunity to create and study the fundamental properties of a new state of strongly interacting matter, the quark-gluon plasma (QGP), where quarks and gluons are deconfined and which is believed to have existed in the early universe \cite{SHURYAK1980,STAR2005,PHENIX2005,ALICE2010}. One of the most important observables for probing the properties of QGP is collective anisotropic flow: the initial spatial anisotropy of the collision zone is converted by pressure gradients into momentum-space anisotropy of the final-state particles \cite{collective_flow,KOLB2001}. The resulting anisotropic collective flow provides important constraints on key QGP properties, including the equation of state, the speed of sound, the specific shear viscosity $\eta/s$, and the bulk viscosity $\zeta/s$ \cite{EOS,speed_of_sound,Viscosity2007,Viscosity2011,shear_bulk_viscosity}.

Relativistic heavy-ion collisions also provide a unique platform for investigating ground-state nuclear structure, particularly nuclear deformation \cite{GG_beta2,BB_Xe_gamma,STAR_Nature,deformation_2024}. Nuclear deformation is of fundamental interest because it reflects the interplay between shell structure and residual valence-nucleon correlations, and is closely connected to nucleosynthesis, nuclear fission, and neutrinoless double-beta decay \cite{RevModPhys.75.121,SCHATZ1998167,nuclear_fission,double_beta_decay}. At high collision energies, the Lorentz-contracted nuclei interact over an extremely short timescale, allowing the collision to probe the instantaneous geometry of the incoming many-body nuclear configuration. As a result, ground-state deformation can leave measurable imprints on the initial collision geometry. Through the subsequent medium evolution, these geometric features can affect flow harmonics $v_n$, mean transverse momentum fluctuations, and flow harmonic correlations, making them sensitive probes of nuclear shape. Recent studies have applied this approach to a wide range of systems, including $^{238}$U, Ru+Ru/Zr+Zr isobars, and $^{129}$Xe, among others, to probe quadrupole deformation and triaxiality, octupole and hexadecapole deformations, as well as possible signatures of nuclear-shape phase transition \cite{STAR_Nature,XHJ_U_beta4,Fortier2025UU,Magdy2023UUFlow,Zhang2026OctupoleU,Zhang_Zr_Ru,Zhao_Zr_Ru,Wang2026IsobarMultiparticle,BB_Xe_gamma,softgamma}.

Despite this progress, existing phenomenological studies have usually constrained nuclear deformation through comparisons among a limited number of prescribed deformation scenarios, or through low-dimensional scans in which only selected deformation parameters are varied. These approaches have been crucial for demonstrating the sensitivity of heavy-ion observables to specific deformation components, such as $\beta_2$, $\gamma$, and $\beta_4$. However, they do not fully characterize how such sensitivities evolve when several deformation parameters are varied simultaneously. The response of a given observable to one parameter may depend on the values of the others, so trends inferred from benchmark configurations or one-dimensional variations need not be directly transferable to a more general deformation configuration. The present study addresses this limitation by using $^{129}$Xe+$^{129}$Xe collisions as a general testing ground for a joint analysis in the multidimensional $(\beta_2,\gamma,\beta_4)$ space, rather than assuming that xenon necessarily contains all these deformation components as established physical inputs. Its main significance is to shift nuclear-shape imaging from single-parameter sensitivity tests toward a multidimensional mapping of how multiparticle observables respond to combined variations of deformation parameters.

The remainder of this paper is organized as follows. Section~\ref{sec:model} presents the model setup and analysis methods, including the modified wounded-quark Glauber initial condition, the iEBE-VISHNU hybrid model for final-state evolution, and the definitions of the multiparticle observables. Section~\ref{sec:results} presents the multidimensional response maps of these observables in the three-dimensional $(\beta_2,\gamma,\beta_4)$ deformation parameter space, compares initial-state estimates with the corresponding final-state results, and examines how the sensitivities of established and newly constructed observables vary across this parameter space. Section~\ref{sec:summary} summarizes the main findings and discusses their implications for future multi-parameter studies of nuclear deformation in relativistic heavy-ion collisions.

\section{Model setup and analysis methods}
\label{sec:model}

\subsection{Initial-state estimator}

Given the high statistics required for calculating multiparticle correlations and the computational cost of a three-dimensional scan in deformation parameter space, we adopt an initial-state approach in the following joint analysis. Nuclear deformation shapes the event-by-event initial geometry through the spatial distribution of nucleons and the random orientation of the incoming nuclei. Initial-state observables therefore provide controlled theoretical proxies that are more directly connected to the nuclear deformation encoded in the collision geometry. They cannot replace experimentally measurable final-state quantities, but they help identify which observables are intrinsically sensitive to nuclear deformation before testing whether that sensitivity survives hydrodynamic evolution, hadronic rescattering, freeze-out effects, and statistical fluctuations.

The initial entropy density profile is generated using a modified Monte Carlo Glauber model with three-quark constituents \cite{WQ1,WQ2,WoundedQuark}. Compared to the standard nucleon Glauber model, the wounded quark model naturally reproduces the approximate linearity between charged-particle multiplicity and the number of wounded quarks, $\mathrm{d}N_{\mathrm{ch}}/\mathrm{d}\eta \propto N_{\mathrm{wq}}$ \cite{NucleonOrQuark,SPS,PHENIX_WQ}.

We use the deformed Woods-Saxon profile to sample the nucleon positions \cite{WS}:
\begin{equation}\label{eq:deformations}
\begin{aligned}
	&\rho(r,\theta,\varphi)\propto\frac{1}{1+\exp\left\{[r-R(\theta,\varphi)]/a\right\}}, \\
	&R(\theta,\varphi)=R_0\{1+\beta_2[\cos\gamma Y_{20}(\theta,\varphi)+\sin\gamma Y_{22}(\theta,\varphi)] \\
		&\qquad\qquad\qquad+\beta_3Y_{30}(\theta,\varphi)+\beta_4Y_{40}(\theta,\varphi)+\cdots\},
\end{aligned}
\end{equation}
where $a$ is the diffuseness parameter, $R_0$ is the nuclear radius, and $Y_{lm}$ are spherical harmonics. The parameters $\beta_2$, $\beta_3$, and $\beta_4$ characterize the quadrupole, octupole, and hexadecapole deformations, while $\gamma$ specifies the degree of triaxiality.

To include subnucleonic degrees of freedom, three constituent quarks are placed around each nucleon according to an exponential radial profile, $\rho_{\mathrm{quarks}}(r) = \frac{r^2}{r_0^3} e^{-r/r_0}$. Their coordinates are then shifted to ensure their center of mass coincides with the sampled nucleon position.
The probability of an inelastic binary collision between a projectile-target quark pair is determined by $p=\exp(-\pi b^2/\sigma_{\mathrm{qq}})$, where $b$ is the transverse separation between the two quarks. A constituent quark is counted as wounded if it undergoes at least one inelastic collision. Following Ref.~\cite{WoundedQuark}, we take $r_0=0.30~\mathrm{fm}$ and $\sigma_{\mathrm{qq}}=13.6~\mathrm{mb}$ at $\sqrt{s_{\mathrm{NN}}}=5.02$ and $5.44~\mathrm{TeV}$.

To account for multiplicity fluctuations, we overlay particle production fluctuations on the wounded-quark sources. Specifically, each wounded quark is assigned an entropy-deposition weight, $n_i$, sampled from a negative binomial distribution \cite{WoundedQuark},
\begin{equation}
  P_{\mathrm{NB}}(n_i|\bar{n},\kappa) = \frac{\Gamma(n_i+\kappa)\bar{n}^{n_i}\kappa^\kappa}{\Gamma(\kappa)n_i!(\bar{n}+\kappa)^{n_i+\kappa}}.
\end{equation}
The parameters $\bar{n}$ and $\kappa$ are taken from Ref.~\cite{WoundedQuark} and globally rescaled to match the experimental multiplicity data. This procedure effectively corresponds to a convolution of the wounded quark distribution with the particle production per source.

Finally, assuming longitudinal boost invariance around mid-rapidity, the initial entropy density profile $s(x,y)$ in the transverse plane is constructed from the coordinates $\vec{x}_i$ of the wounded quarks and their corresponding weights $n_i$ \cite{GaussianSmearing}:
\begin{equation}\label{eq:entropy_density}
    s(x,y)\propto\frac{1}{2\pi\sigma^2}\sum_{i=1}^{N_{\mathrm{wq}}}n_i\cdot \exp\left[-\frac{(\vec x-\vec x_i)^2}{2\sigma^2}\right].
\end{equation}
Here we use a Gaussian smearing width $\sigma=0.55~\mathrm{fm}$.

Motivated by the approximate hydrodynamic response between the initial-state anisotropies and final-state flow harmonics, the eccentricity vectors $\mathcal E_n$ are used as geometric estimators for flow vectors $V_n$ \cite{linear_response}. The $n$-th eccentricity is computed by
\begin{equation}\label{eq:ecc}
	\mathcal E_n\equiv\varepsilon_n\exp(in\Psi_n)=-\frac{\int\mathrm dx\mathrm dy s(x,y)\cdot(x+iy)^n}{\int\mathrm dx\mathrm dy s(x,y)\cdot|x+iy|^n}.
\end{equation}

Initial-state estimates of the mean transverse momentum $[p_{\mathrm{T}}]$ typically rely on either geometric or thermodynamic proxies \cite{Schenke2020, Giacalone2020}. In this work, we follow Ref.~\cite{Giacalone2020} and estimate $[p_{\mathrm{T}}]$ from the initial energy per unit entropy, $E/S$. At sufficiently high temperature, the equation of state of hot QCD matter implies the approximate relation between the local energy density and entropy density, $\epsilon \propto s^{4/3}$, where $s$ is obtained in Eq.~(\ref{eq:entropy_density}). This leads to:
\begin{equation}
  [p_{\mathrm{T}}] \propto \frac{E}{S} \propto \frac{\int \mathrm{d}x\mathrm{d}y \, s^{4/3}(x,y)}{\int \mathrm{d}x\mathrm{d}y \, s(x,y)}.
\end{equation}

Within this initial-state framework, we obtain event-by-event $N_{\mathrm{ch}}$, $\mathcal E_n$, and $[p_{\mathrm{T}}]$, thereby enabling the estimation of multiparticle observables in specific centrality bins. We then perform a three-dimensional scan of the deformation-parameter space by uniformly sampling 500 parameter sets spanning $\beta_2 \in [0.0, 0.4]$, $\gamma \in [0^\circ, 60^\circ]$, and $\beta_4 \in [0.0, 0.2]$. For each parameter set, $4\times10^5$ events are simulated to achieve sufficient statistical precision.

\subsection{iEBE-VISHNU hybrid model}

Although the initial-state estimator provides a cleaner mapping to the geometric effects of nuclear deformation, the corresponding dependence of final-state observables may be altered by hydrodynamic response, hadronization, and hadronic rescattering. We therefore perform full hybrid simulations to examine whether the resulting deformation sensitivities survive the subsequent dynamical evolution. For this purpose, we employ the state-of-the-art iEBE-VISHNU model \cite{iEBE-VISHNU,SHC_frzout,Moreland} to simulate $^{129}$Xe+$^{129}$Xe collisions at $\sqrt{s_{\mathrm{NN}}}=5.44~\mathrm{TeV}$, with $^{208}$Pb+$^{208}$Pb collisions at $\sqrt{s_{\mathrm{NN}}}=5.02~\mathrm{TeV}$ serving as a spherical reference.

In the iEBE-VISHNU hybrid model, the initial condition is generated by TRENTo\cite{Trento}, followed by the (2+1)-dimensional viscous hydrodynamic evolution with VISH2+1\cite{SHC_Hydro,SHC_Hydro2}, Cooper-Frye particlization near the transition temperature\cite{SHC_frzout}, and a hadronic afterburner described by UrQMD\cite{UrQMD1,UrQMD2}. The specific parameters are taken to be consistent with Ref.~\cite{softgamma} for $^{129}$Xe+$^{129}$Xe and Ref.~\cite{Pb_parameters} for $^{208}$Pb+$^{208}$Pb collisions. Since central collisions maximize the sensitivity of multiparticle correlations to nuclear deformation, we focus on the 0--5\% centrality class, defined by an initial-entropy trigger in TRENTo.

\begin{table}[b]
\caption{\label{table:WS}Deformation parameter sets used for the Woods-Saxon density distributions of $^{129}\mathrm{Xe}$. For all configurations, the nuclear radius and diffuseness parameter are fixed to $R_0=5.36~\mathrm{fm}$ and $a=0.590~\mathrm{fm}$ \cite{R0_a_set}.}
\begin{ruledtabular}
\begin{tabular}{cccc}
$\beta_2$ & $\gamma$ & $\beta_4$ & Remark \\
\colrule
0.17 & $30^\circ$ & 0.0 & baseline \\
0.17 & $30^\circ$ & 0.1 & non-zero $\beta_4$ \\
0.27 & $30^\circ$ & 0.0 & enhanced $\beta_2$ \\
0.27 & $30^\circ$ & 0.1 & enhanced $\beta_2$ and non-zero $\beta_4$ \\
0.17 & $0^\circ$  & 0.0 & prolate limit \\
0.17 & $60^\circ$ & 0.0 & oblate limit \\
\end{tabular}
\end{ruledtabular}
\end{table}

Because a full three-dimensional scan with the iEBE-VISHNU hybrid model is computationally prohibitive, we restrict the final-state calculations to a set of representative deformation configurations. Six Woods-Saxon density distributions are implemented for $^{129}$Xe, as listed in Table~\ref{table:WS}. We choose a baseline triaxial configuration with $\beta_2=0.17$, $\gamma=30^\circ$, and $\beta_4=0.0$, following Ref.~\cite{softgamma}. To investigate the sensitivities to quadrupole and hexadecapole deformations, we vary the deformation parameters by introducing either an enhanced $\beta_2$, a non-zero $\beta_4$, or a combination of both. To isolate the effect of triaxiality, we vary $\gamma$ from $0^\circ$ to $60^\circ$, corresponding to the prolate and oblate limits, while fixing $\beta_2=0.17$ and $\beta_4=0.0$.

For each deformation configuration in $^{129}$Xe+$^{129}$Xe collisions, as well as the spherical $^{208}$Pb+$^{208}$Pb system ($R_0=6.62~\mathrm{fm},~a=0.546~\mathrm{fm}$), we simulate about $10^6$ events for the 0--5\% central collision class. The multiparticle observables are evaluated with the standard Q-cumulant method \cite{Q_cumulant,Q_cumulant_2}, using charged particles in the kinematic range $0.2 < p_{\mathrm{T}} < 3.0~\mathrm{GeV}/c$ and $|\eta| < 0.8$ \cite{kinetic_cut}. The corresponding results and discussions are presented in the following section~\ref{sec:results}.

\subsection{Multiparticle observables}
\label{subsec:obs}

The azimuthal angle distribution of final-state particles in momentum space can be expressed as a Fourier series
\begin{equation}
	\frac{\mathrm{d} N}{\mathrm d\varphi}\propto 1+\sum_{n=1}^\infty 2v_n\cos\left[n(\varphi-\Phi_n)\right],
\end{equation}
and defines the complex flow vectors $V_n=v_n e^{in\Phi_n}$. The flow observables can be calculated by the $m$-particle azimuthal correlation \cite{Q_cumulant_2}:
\begin{equation}
    \langle m\rangle_{n_1,n_2,\cdots,n_m}
    \equiv
    \left\langle e^{i(n_1\varphi_{k_1}+n_2\varphi_{k_2}+\cdots+n_m\varphi_{k_m})}\right\rangle,
\end{equation}
where $\langle\cdots\rangle$ denotes the average over all particles of interest (POI) in a single event. The two-particle $n$th-order flow harmonic is then defined as \cite{Q_cumulant}
\begin{equation}
    v_n\{2\}^2 \equiv \langle\langle 2\rangle_{n,-n}\rangle,
\end{equation}
where $\langle\langle\cdots\rangle\rangle$ denotes the subsequent average of $\langle\cdots\rangle$ over an ensemble of events. The corresponding four-particle moment is given by \cite{Q_cumulant}
\begin{equation}
    \langle v_n^4\rangle \equiv \langle\langle 4\rangle_{n,n,-n,-n}\rangle.
\end{equation}
To characterize the nonlinear coupling between different flow harmonics, we also consider the three-particle asymmetric cumulant \cite{asymmetric_c}
\begin{equation}\label{eq:ac_nm}
    \mathrm{ac}_{nm}\equiv\langle\langle3\rangle_{n,m,-(n+m)}\rangle.
\end{equation}
In the absence of non-flow effects, it can be written as
\begin{equation}\label{eq:ac_flow}
    \mathrm{ac}_{nm}=\left\langle v_n v_m v_{n+m}\cos\left(n\Phi_n+m\Phi_m-(n+m)\Phi_{n+m}\right)\right\rangle.
\end{equation}

A widely used observable is the correlation between $v_n^2$ and mean transverse momentum $[p_{\mathrm{T}}]$, evaluated by the Pearson correlation coefficient \cite{rho2}:
\begin{equation}\label{eq:rho_n}
    \rho_n \equiv \rho(v_n\{2\}^2,[p_{\mathrm{T}}])
    = \frac{\mathrm{cov}(\langle2\rangle_{n,-n},[p_{\mathrm{T}}])}{\sqrt{\mathrm{var}(\langle2\rangle_{n,-n})\mathrm{var}([p_{\mathrm{T}}])}},
\end{equation}
where $\mathrm{cov}(\cdots)$ and $\mathrm{var}(\cdots)$ denote the covariance and variance over events in a given centrality class. The $n=2$ case, $\rho_2$, is sensitive to quadrupole deformation and has been broadly used to probe nuclear shape in central heavy-ion collisions \cite{GG_beta2,BB_Xe_gamma,STAR_Nature}. Nevertheless, observables involving higher moments or mixed-harmonic couplings can access more complex correlations, motivating their use as complementary probes of nuclear deformation.

In particular, the four-particle nonlinear response coefficient $\chi_{4,22}$ was introduced to isolate the $\beta_4$ signal in $^{238}$U, which is otherwise typically overwhelmed by the large $\beta_2$ background in conventional observables \cite{XHJ_U_beta4}. This observable is defined as \cite{asymmetric_c}
\begin{equation}\label{eq:chi422}
	\chi_{4,22}\equiv\frac{v_4\{\Phi_2\}}{\langle v_2^4\rangle^{1/2}}=\frac{\langle\langle3\rangle_{2,2,-4}\rangle}{\langle\langle 4\rangle_{2,2,-2,-2}\rangle}.
\end{equation}

To access correlations among different harmonic components and their correlation to mean transverse momentum, we further consider the Pearson correlation coefficient between different flow harmonics,
\begin{equation}\label{eq:rho_nm}
\begin{aligned}
	\rho_{nm}&\equiv\rho(v_n^2\{2\},v_m^2\{2\}) \\
	&=\cfrac{\mathrm{cov}(\langle2\rangle_{n,-n},\langle2\rangle_{m,-m})}{\sqrt{\mathrm{var}(\langle2\rangle_{n,-n})\mathrm{var}(\langle2\rangle_{m,-m})}},
\end{aligned}
\end{equation}
as well as the Pearson correlation coefficient between the asymmetric cumulant and $[p_{\mathrm{T}}]$,
\begin{equation}\label{eq:rho_nmnm}
\begin{aligned}
	\rho_{nm(n+m)}&\equiv\rho(\mathrm{ac}_{nm},[p_{\mathrm{T}}]) \\
	&=\cfrac{\mathrm{cov}(\langle3\rangle_{n,m,-(n+m)},[p_{\mathrm{T}}])}{\sqrt{\mathrm{var}(\langle3\rangle_{n,m,-(n+m)})\mathrm{var}([p_{\mathrm{T}}])}}.
\end{aligned}
\end{equation}

In the initial-state framework, we construct geometric counterparts of the above observables using the eccentricity vectors $\mathcal E_n=\varepsilon_n e^{in\Psi_n}$. The corresponding initial-state estimators are defined as
\begin{equation}
    \rho_n^{(\mathrm{init})}=\rho(\varepsilon_n^2,[p_{\mathrm{T}}]), \quad
    \rho_{nm}^{(\mathrm{init})}=\rho(\varepsilon_n^2,\varepsilon_m^2).
\end{equation}

Guided by Eq.~(\ref{eq:ac_flow}), the asymmetric cumulant in the initial-state framework is estimated as
\begin{equation}
    \mathrm{ac}_{nm}^{(\mathrm{init})}=\left\langle
    \varepsilon_n \varepsilon_m \varepsilon_{n+m}\cos\left(n\Psi_n+m\Psi_m-(n+m)\Psi_{n+m}\right)\right\rangle,
\end{equation}
which leads to the nonlinear response coefficient
\begin{equation}
	\chi_{4,22}^{(\mathrm{init})}=\frac{\mathrm{ac}_{22}^{(\mathrm{init})}}{\langle\varepsilon_2^4\rangle}.
\end{equation}
Similarly, the Pearson correlation coefficient between the asymmetric cumulant and $[p_{\mathrm{T}}]$ is defined as
\begin{equation}
    \rho_{nm(n+m)}^{(\mathrm{init})}=\rho(\mathrm{ac}_{nm}^{(\mathrm{init})},[p_{\mathrm{T}}]).
\end{equation}
These quantities serve as initial-state proxies for the corresponding final-state observables. In the following analysis, the deformation dependence of each observable $\mathcal{O}$ is presented through its ratio between $^{129}$Xe+$^{129}$Xe and a spherical $^{208}$Pb+$^{208}$Pb reference system,
\begin{equation}
    R(\mathcal O) \equiv
    \frac{\mathcal O_{^{129}\mathrm{Xe}+^{129}\mathrm{Xe}}(\beta_2,\gamma,\beta_4)}{\mathcal O_{^{208}\mathrm{Pb}+^{208}\mathrm{Pb}}}.
\end{equation}
This ratio representation helps reduce the impact of common model ingredients and analysis choices, and highlights the deformation-driven variations specific to the $^{129}$Xe system.

\section{Results and discussion}
\label{sec:results}

To clearly visualize the multidimensional dependence of multiparticle observables on three deformation parameters, the corresponding three-dimensional (3D) distributions are presented in the following as two-dimensional (2D) cross-sectional contour plots: specifically, the slice on the $(\beta_2, \gamma)$ plane with a fixed $\beta_4=0.0$, and another slice on the $(\beta_2, \beta_4)$ plane with a fixed $\gamma=30^\circ$.

\subsection{\texorpdfstring{$v_2^2-[p_\mathrm{T}]$}{v2 squared--pT} correlation}

\begin{figure}[t]
	\centering
	\includegraphics[width=0.45\textwidth]{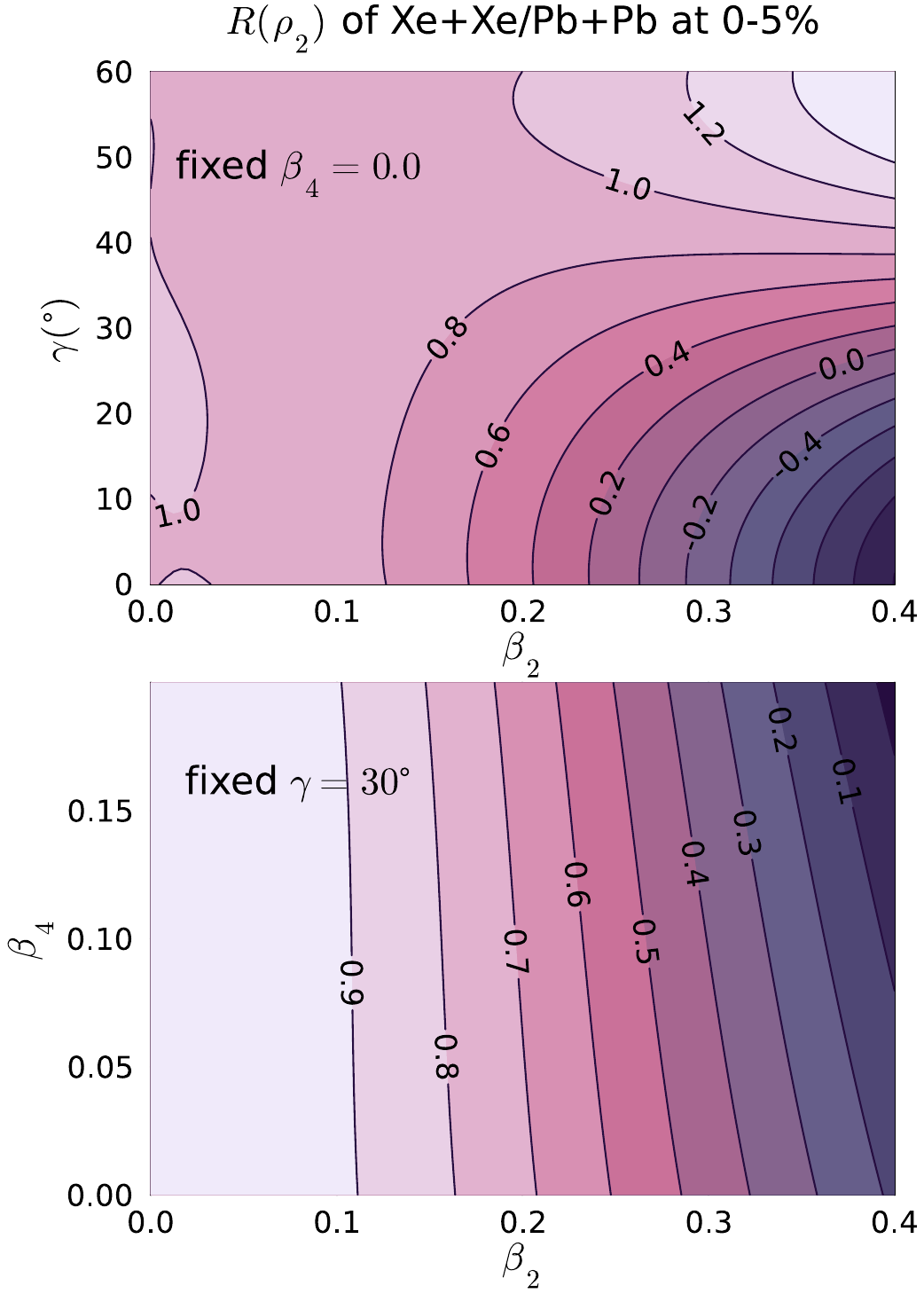}
	\caption{Contour plots of the initial-state estimator for the ratio $R(\rho_{2})$ in 0--5\% central $^{129}$Xe+$^{129}$Xe at $\sqrt{s_{\mathrm{NN}}}=5.44~\mathrm{TeV}$ over $^{208}$Pb+$^{208}$Pb collisions at $\sqrt{s_{\mathrm{NN}}}=5.02~\mathrm{TeV}$. The upper and lower panels show the slices on $(\beta_2, \gamma)$ plane at $\beta_4 = 0.0$ and $(\beta_2, \beta_4)$ plane at $\gamma = 30^\circ$, respectively.}
	\label{fig:rho2_contourf}
\end{figure}

We begin with the well-established deformation probe $\rho_2$ in the 3D joint analysis. Fig.~\ref{fig:rho2_contourf} displays the 2D slices of $R(\rho_2)$ obtained from the initial-state model in the parameter space $(\beta_2, \gamma, \beta_4)$ for 0--5\% ultra-central collisions.

In the upper panel, corresponding to $\beta_4 = 0.0$, $\rho_2$ displays a pronounced sensitivity to both $\beta_2$ and $\gamma$. At fixed $\gamma$, $R(\rho_2)$ generally decreases as $\beta_2$ increases, and the ratio can even change from positive to negative values at large $\beta_2$ for small $\gamma$. At fixed $\beta_2$, $R(\rho_2)$ increases monotonically with $\gamma$, with the enhancement becoming much more visible at larger $\beta_2$. This behavior is expected because $\gamma$ characterizes the triaxiality of the quadrupole deformation, so changing $\gamma$ has no physical effect when $\beta_2=0$. Conversely, the dependence on $\beta_2$ is strongest near the prolate limit $\gamma=0^\circ$, where increasing $\beta_2$ leads to a rapid suppression of $R(\rho_2)$.

In the lower panel with $\gamma = 30^\circ$, $R(\rho_2)$ decreases monotonically with increasing $\beta_2$ over the whole $\beta_4$ range. The contour lines are nearly perpendicular to the $\beta_2$ axis, showing that changing $\beta_4$ from 0 to 0.2 produces only a weak variation compared with the dominant $\beta_2$ dependence. This indicates that $\rho_2$ is primarily controlled by the quadrupole deformation and is largely insensitive to $\beta_4$.

It is worth noting that when averaging over a uniform distribution of $\gamma\in[0^\circ,60^\circ]$ in this 3D parameter space, the resulting 2D projection on the $(\beta_2, \beta_4)$ plane is highly similar to the case with a fixed $\gamma=30^\circ$. This observation indirectly supports the findings in Ref.~\cite{softgamma} that $\rho_2$ alone is insufficient to distinguish between rigid triaxial deformation and a $\gamma$-soft nuclear structure.

To quantify the global sensitivity of $\rho_2$ to each deformation parameter in this 3D space, the distance correlation (dCor) is introduced as the dependence measure, which can capture the nonlinear couplings among variables \cite{dCor2007,GSA2013}. For two sampled variables $X=\{x_i\}_{i=1}^{N}$ and $Y=\{y_i\}_{i=1}^{N}$, we first construct the pairwise distance matrices $a_{ij}=|x_i-x_j|$ and $b_{ij}=|y_i-y_j|$, and double-center them as
\begin{equation}
    A_{ij}=a_{ij}-\bar a_{i\cdot}-\bar a_{\cdot j}+\bar a_{\cdot\cdot},\quad
    B_{ij}=b_{ij}-\bar b_{i\cdot}-\bar b_{\cdot j}+\bar b_{\cdot\cdot}.
\end{equation}
Here $\bar a_{i\cdot}=\frac1N\sum_j a_{ij}$, $\bar a_{\cdot j}=\frac1N\sum_i a_{ij}$, and $\bar a_{\cdot\cdot}=\frac1{N^2}\sum_{i,j}a_{ij}$ denote the row mean, column mean, and grand mean of $a_{ij}$, respectively, with analogous definitions for the $b$ matrix.

The sample distance correlation is then defined as \cite{dCor2007}
\begin{equation}\label{eq:dCor}
    \mathrm{dCor}(X,Y)=
    \left[
    \frac{\mathcal V_N^2(X,Y)}
    {\sqrt{\mathcal V_N^2(X,X)\mathcal V_N^2(Y,Y)}}
    \right]^{1/2},
\end{equation}
with
\begin{equation}
    \mathcal V_N^2(X,Y)=\frac{1}{N^2}\sum_{i,j=1}^{N}A_{ij}B_{ij}.
\end{equation}

Using this definition, we obtain $\mathrm{dCor}(\rho_2,\beta_2)=0.606$ and $\mathrm{dCor}(\rho_2,\gamma)=0.611$, while $\mathrm{dCor}(\rho_2,\beta_4)$ is merely $0.044$. This quantitative analysis confirms that $\rho_2$ is strongly sensitive to quadrupole and triaxial deformations, but remains largely insensitive to hexadecapole deformation.

\begin{figure}[t]
	\centering
	\includegraphics[width=0.45\textwidth]{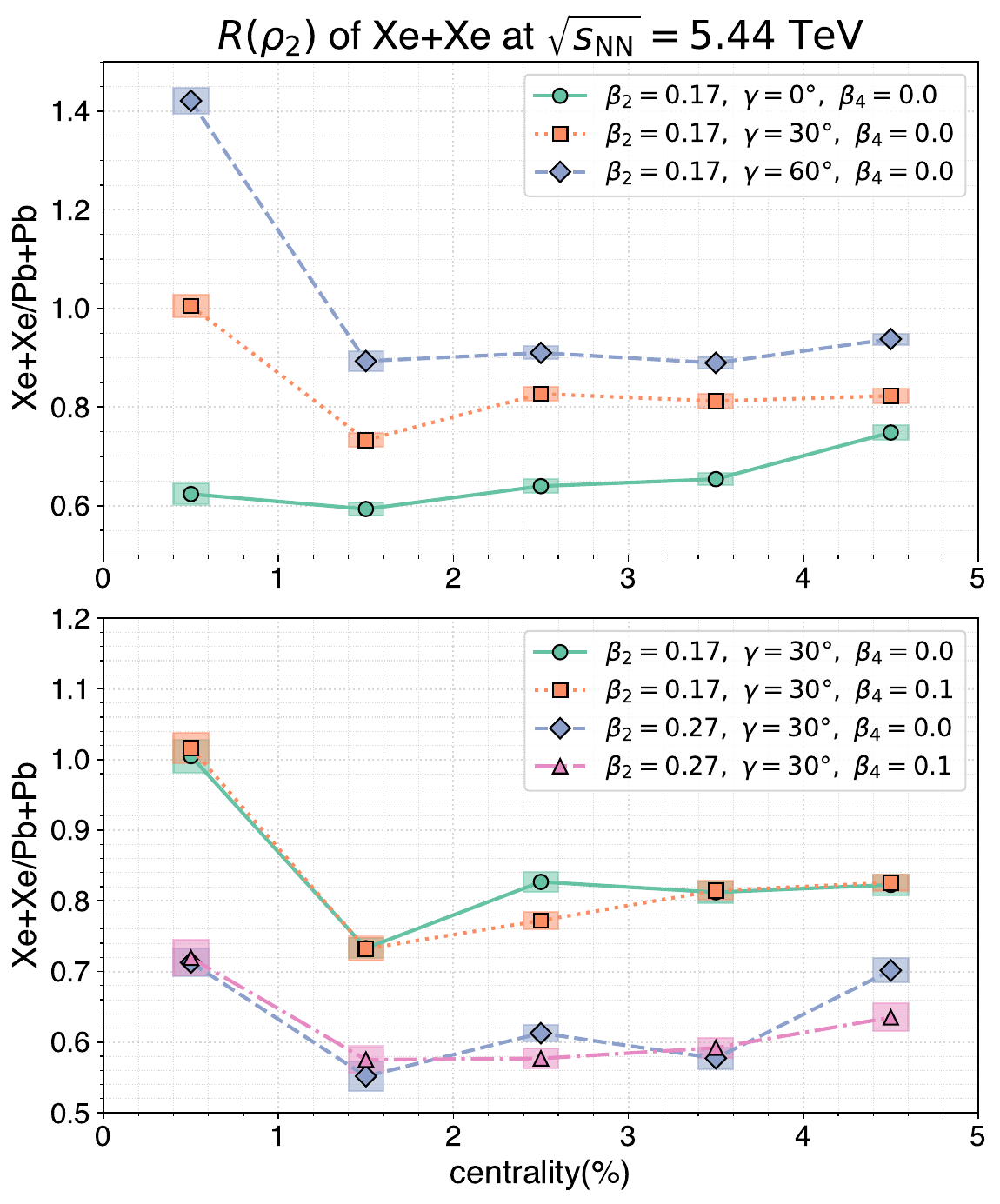}
	\caption{Centrality dependence of $R(\rho_{2})$ calculated by iEBE-VISHNU model. The upper panel shows the effect of varying $\gamma$ at fixed $\beta_2 = 0.17$ and $\beta_4 = 0.0$, while the lower panel compares configurations with different $\beta_2$ and $\beta_4$ values at fixed $\gamma = 30^{\circ}$.}
	\label{fig:rho2}
\end{figure}

To examine whether these initial-state geometric effects survive after full dynamical evolution, the iEBE-VISHNU simulation results are presented in Fig.~\ref{fig:rho2}. In the upper panel, a pronounced increase of $R(\rho_2)$ from $\gamma=0^\circ$ to $60^\circ$ is observed. In the lower panel with fixed $\gamma=30^\circ$, varying $\beta_2$ from 0.17 to 0.27 leads to prominent suppression of $R(\rho_2)$, whereas changing $\beta_4$ from 0.0 to 0.1 has a negligible effect, as indicated by the nearly overlapping curves. These final-state results are consistent with the corresponding initial-state estimator constructed from the collision geometry, supporting $\rho_2$ as a robust probe of $\beta_2$ and $\gamma$ in nuclear structure.

\subsection{Nonlinear response coefficient \texorpdfstring{$\chi_{4,22}$}{chi422}}

We next turn to the nonlinear response coefficient $\chi_{4,22}$. In the final state, the fourth-order flow vector contains both a linear component and a nonlinear contribution driven by elliptic flow \cite{asymmetric_c},
\begin{equation}
    V_4 = V_4^{L}+\chi_{4,22}V_2^2 .
\end{equation}
Although $\chi_{4,22}$ is a final-state response coefficient, its sensitivity to nuclear deformation can be guided by the underlying geometric correlations in the initial state. In particular, the fourth-order spatial anisotropy is not independent of the elliptic geometry, and correlations between $\mathcal E_4$ and $\mathcal E_2^2$ provide the geometric analogue of the nonlinear coupling between $V_4$ and $V_2^2$ \cite{Nonlinearities}. Therefore, we use $\chi_{4,22}^{(\mathrm{init})}$ as a geometric baseline and compare it with the iEBE-VISHNU results.

\begin{figure}[t]
	\centering
	\includegraphics[width=0.45\textwidth]{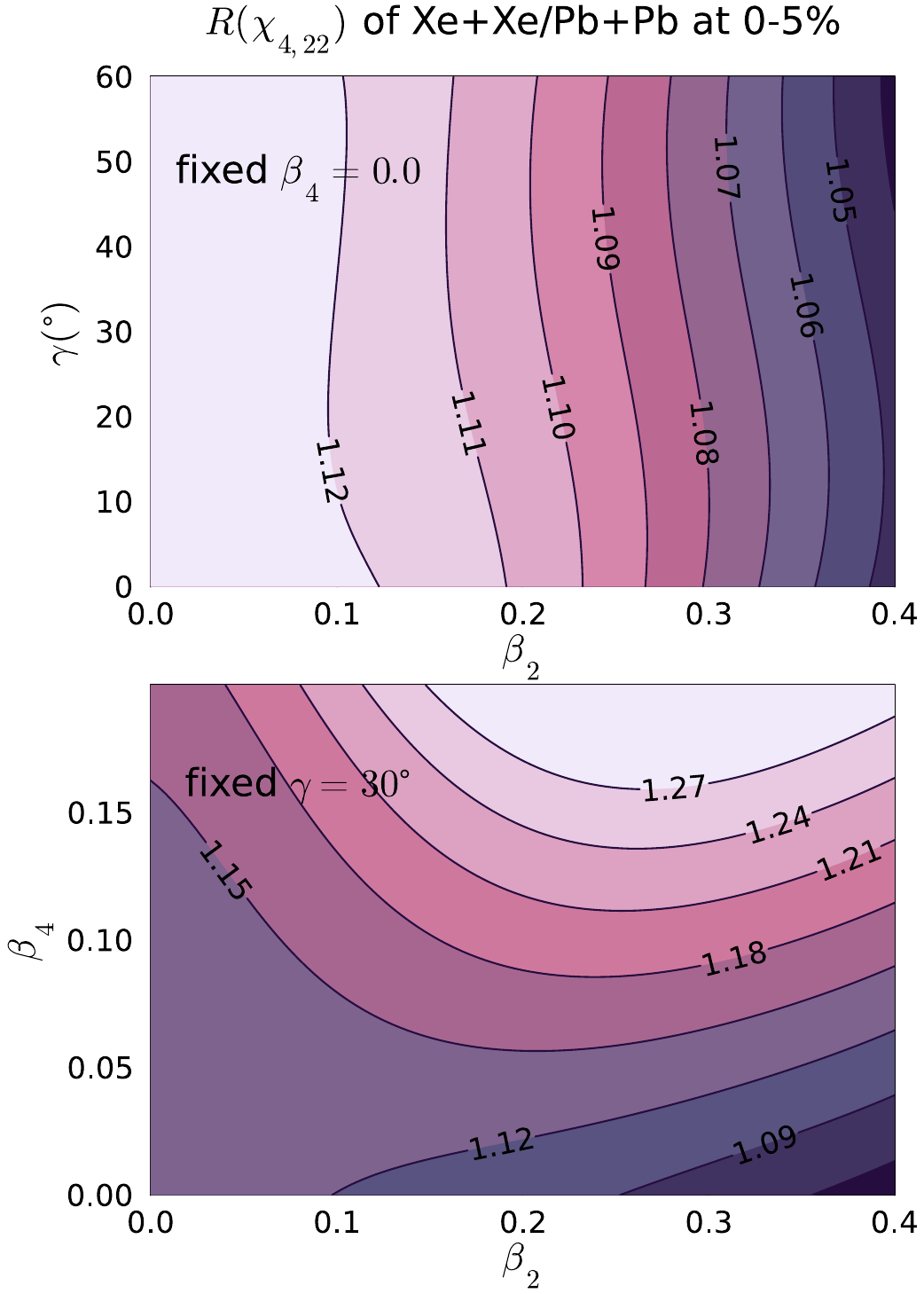}
	\caption{Contour plots of the initial-state estimator for the ratio $R(\chi_{4,22})$ in 0--5\% central $^{129}$Xe+$^{129}$Xe at $\sqrt{s_{\mathrm{NN}}}=5.44~\mathrm{TeV}$ over $^{208}$Pb+$^{208}$Pb collisions at $\sqrt{s_{\mathrm{NN}}}=5.02~\mathrm{TeV}$. The upper and lower panels show the slices on $(\beta_2, \gamma)$ plane at $\beta_4 = 0.0$ and $(\beta_2, \beta_4)$ plane at $\gamma = 30^\circ$, respectively.}
	\label{fig:chi422_contourf}
\end{figure}

\begin{figure}[t]
	\centering
	\includegraphics[width=0.45\textwidth]{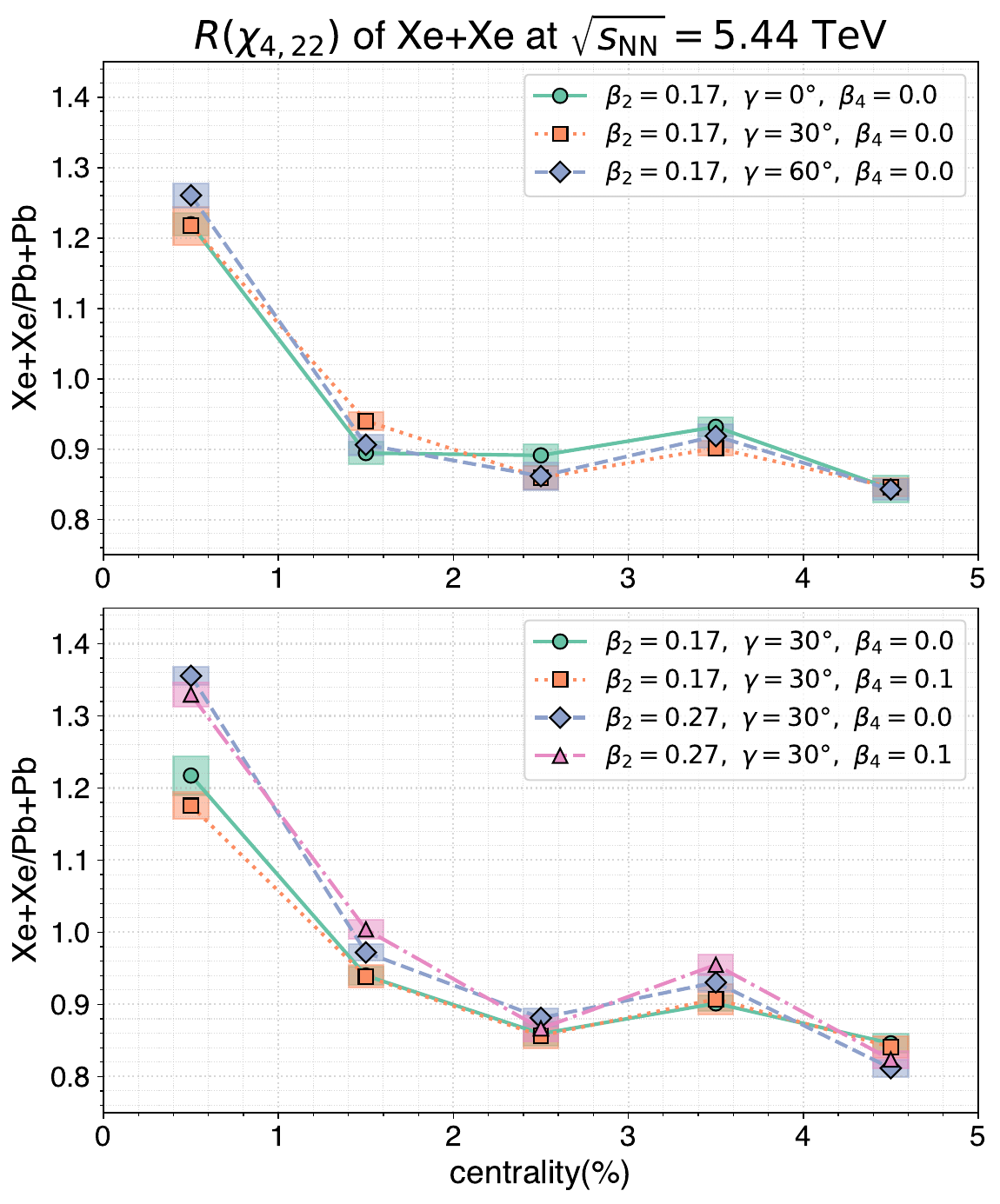}
	\caption{Centrality dependence of $R(\chi_{4,22})$ calculated by iEBE-VISHNU model. The upper panel shows the effect of varying $\gamma$ at fixed $\beta_2 = 0.17$ and $\beta_4 = 0.0$, while the lower panel compares configurations with different $\beta_2$ and $\beta_4$ values at fixed $\gamma = 30^{\circ}$.}
	\label{fig:chi422}
\end{figure}

Fig.~\ref{fig:chi422_contourf} presents the 2D slices of the initial-state estimator for $\chi_{4,22}$ in 0--5\% ultra-central collisions. In the upper panel at $\beta_4=0.0$, $R(\chi_{4,22})$ changes only mildly across the $(\beta_2,\gamma)$ plane. The contours show that $R(\chi_{4,22})$ decreases by only about 0.07 as $\beta_2$ increases from 0.0 to 0.4, while the dependence on $\gamma$ mainly appears as a moderate bending of the contour lines. This indicates that, in the absence of hexadecapole deformation, $\chi_{4,22}^{(\mathrm{init})}$ is only weakly affected by quadrupole and triaxial deformations.

In contrast, the lower panel exhibits a pronounced dependence on $\beta_4$. Increasing $\beta_4$ from 0 to 0.2 enhances $R(\chi_{4,22})$ by roughly 0.15--0.2 over a broad range of $\beta_2$. The tilted contour lines further indicate that the sensitivity to $\beta_4$ is not limited to the large-$\beta_2$ region, although its magnitude varies with $\beta_2$. Therefore, the initial-state nonlinear coefficient $\chi_{4,22}^{(\mathrm{init})}$ can serve as a robust and ideal probe for the hexadecapole structure.

We also evaluate the quantitative measures of global sensitivities for $\chi_{4,22}^{(\mathrm{init})}$ by the distance correlation defined in Eq.~(\ref{eq:dCor}). The result is $\mathrm{dCor}(\chi_{4,22},\beta_4)=0.792$, compared with substantially smaller $\mathrm{dCor}(\chi_{4,22},\beta_2)=0.319$ and $\mathrm{dCor}(\chi_{4,22},\gamma)=0.289$. This further confirms that at the initial-state level, $\chi_{4,22}$ is predominantly governed by the hexadecapole deformation.

Fig.~\ref{fig:chi422} presents the centrality dependence of the final-state $R(\chi_{4,22})$ calculated with the iEBE-VISHNU hybrid model. In the upper panel, the curves for different $\gamma$ nearly overlap across the 0--5\% centrality range, consistent with the weak $\gamma$ dependence observed in the initial-state estimator.

However, the lower panel deviates from the initial-state geometric predictions regarding $\beta_2$ and $\beta_4$. The final-state results exhibit a visible splitting when $\beta_2$ is increased from 0.17 to 0.27, whereas the difference between $\beta_4=0.0$ and $\beta_4=0.1$ at fixed $\beta_2$ remains indistinguishable within statistical uncertainties. This indicates that, after the full dynamical evolution, $\chi_{4,22}$ in $^{129}$Xe+$^{129}$Xe collisions is dominated by $\beta_2$ variation, while its resolving power for $\beta_4$ is largely diminished.

While Ref.~\cite{XHJ_U_beta4} demonstrated that $\chi_{4,22}$ can isolate the $\beta_4$ signal in $^{238}$U+$^{238}$U collisions, our results suggest that in a smaller system like $^{129}$Xe+$^{129}$Xe, the hexadecapole deformation contribution may be more strongly suppressed by enhanced viscous damping, or obscured by relatively larger background fluctuations during the subsequent dynamical evolution. Thus, the effectiveness of $\chi_{4,22}$ as a probe of $\beta_4$ appears to be system dependent and requires further investigation.

\subsection{\texorpdfstring{$\mathrm{ac}_{nm}-[p_\mathrm{T}]$}{acnm--pT} correlation}

\begin{figure}[t]
	\centering
	\includegraphics[width=0.45\textwidth]{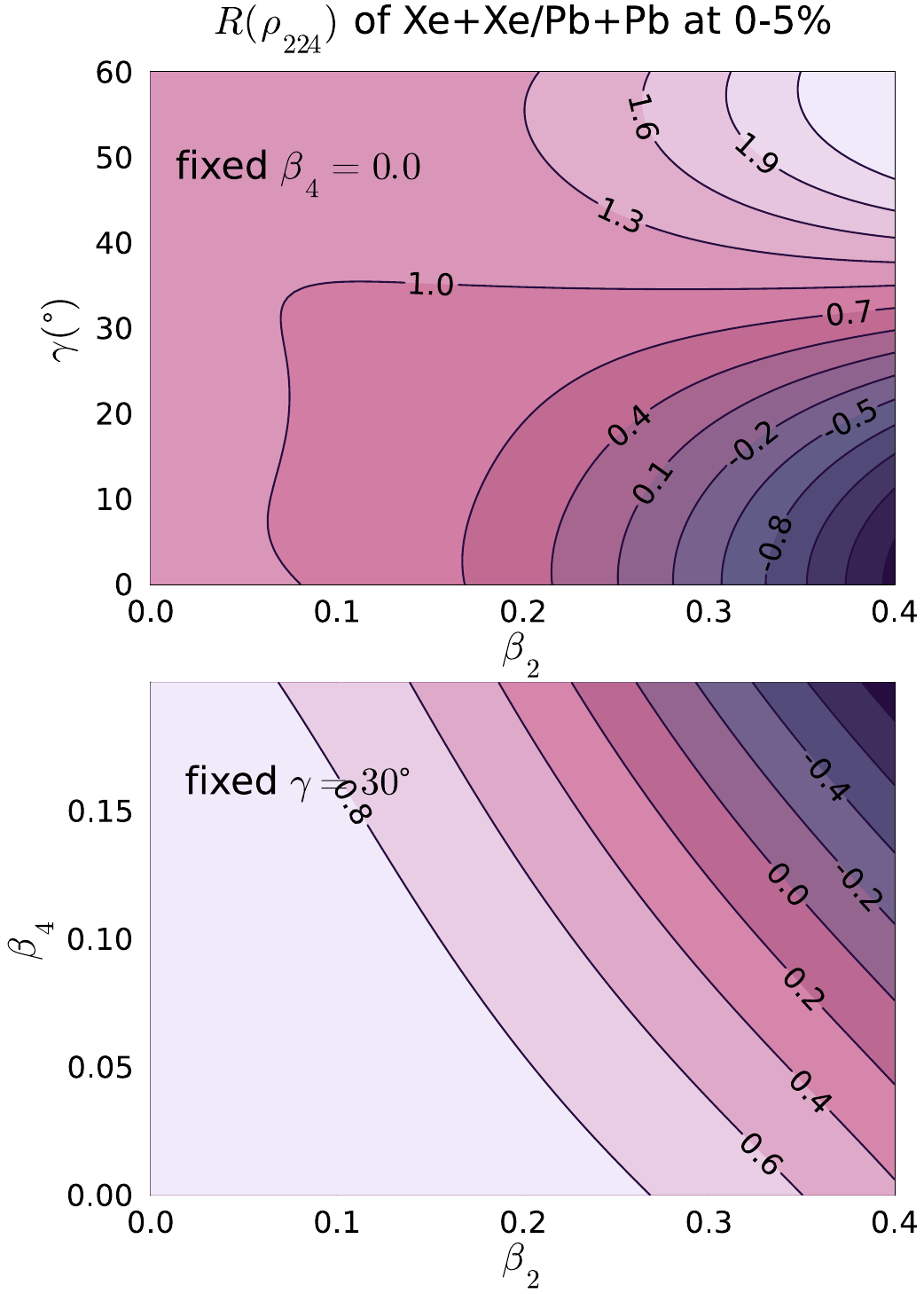}
	\caption{Contour plots of the initial-state estimator for the ratio $R(\rho_{224})$ in 0--5\% central $^{129}$Xe+$^{129}$Xe at $\sqrt{s_{\mathrm{NN}}}=5.44~\mathrm{TeV}$ over $^{208}$Pb+$^{208}$Pb collisions at $\sqrt{s_{\mathrm{NN}}}=5.02~\mathrm{TeV}$. The upper and lower panels show the slices on $(\beta_2, \gamma)$ plane at $\beta_4 = 0.0$ and $(\beta_2, \beta_4)$ plane at $\gamma = 30^\circ$, respectively.}
	\label{fig:rho224_contourf}
\end{figure}

\begin{figure}[t]
	\centering
	\includegraphics[width=0.45\textwidth]{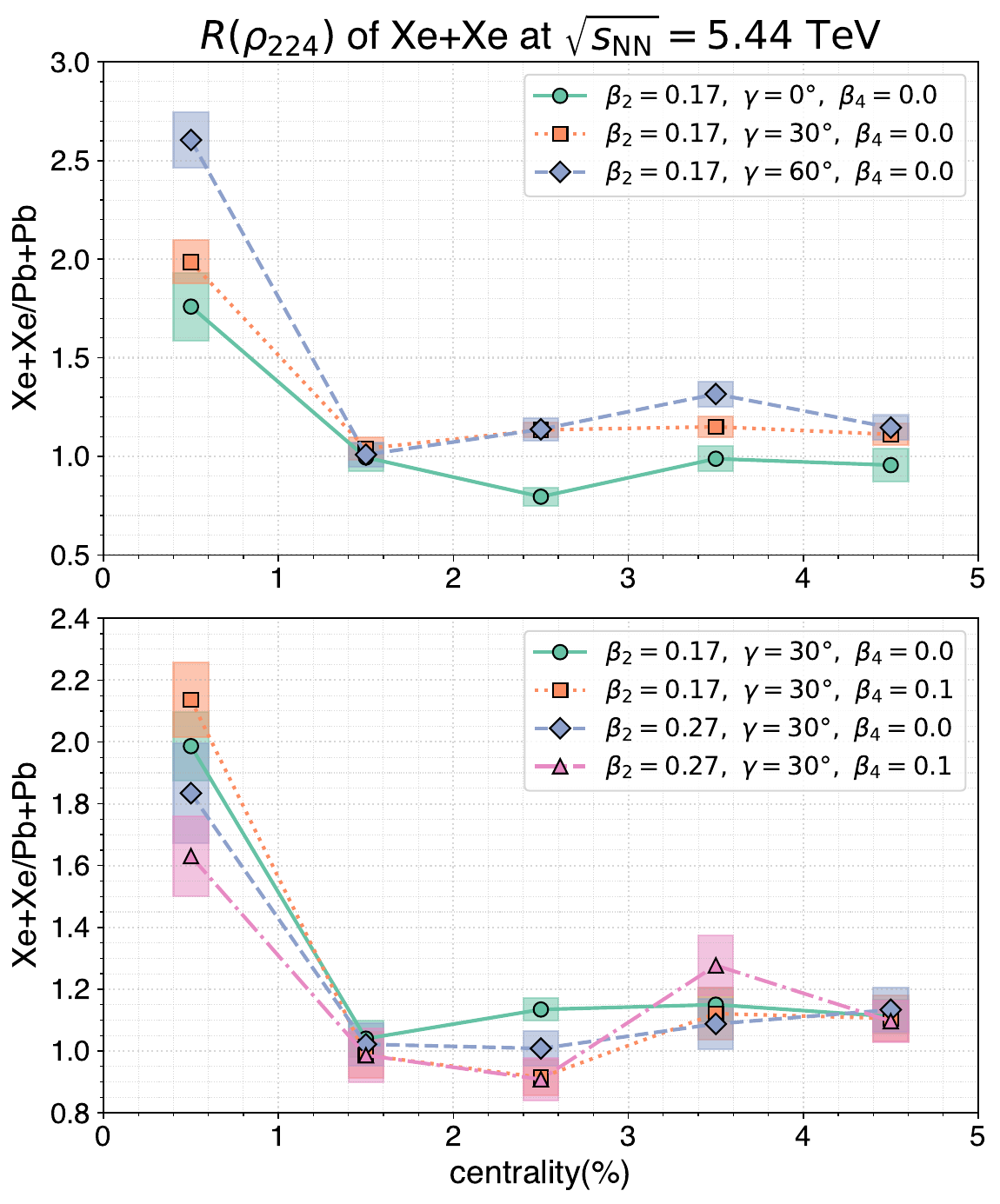}
	\caption{Centrality dependence of $R(\rho_{224})$ calculated by iEBE-VISHNU model. The upper panel shows the effect of varying $\gamma$ at fixed $\beta_2 = 0.17$ and $\beta_4 = 0.0$, while the lower panel compares configurations with different $\beta_2$ and $\beta_4$ values at fixed $\gamma = 30^{\circ}$.}
	\label{fig:rho224}
\end{figure}

In this section, we investigate the multidimensional dependence of the four-particle observable $\rho_{nm(n+m)}$, as introduced in Eq.~(\ref{eq:rho_nmnm}), on the nuclear deformation parameters. Fig.~\ref{fig:rho224_contourf} presents the 2D contour slices of the initial-state estimator for $R(\rho_{224})$. In the upper panel, $R(\rho_{224})$ exhibits pronounced sensitivities to both $\beta_2$ and $\gamma$. Similar to $R(\rho_2)$ in Fig.~\ref{fig:rho2_contourf}, the $\gamma$ dependence becomes particularly visible at larger $\beta_2$, while the observable changes strongly across the full $\beta_2$ range.

A notable difference appears in the lower $(\beta_2,\beta_4)$ panel. The contours are no longer approximately vertical, but exhibit an obvious $\beta_4$-dependent distortion, indicating a non-negligible hexadecapole effect. This $\beta_4$ dependence becomes more pronounced at larger $\beta_2$, suggesting a combined influence of quadrupole and hexadecapole deformations on $\rho_{224}^{(\mathrm{init})}$. The distance-correlation analysis supports this observation: $\mathrm{dCor}(\rho_{224},\beta_2)=0.563$ and $\mathrm{dCor}(\rho_{224},\gamma)=0.643$ remain the dominant sensitivities, while $\mathrm{dCor}(\rho_{224},\beta_4)$ reaches 0.188.

The iEBE-VISHNU simulation results of $\rho_{224}$ are presented in Fig.~\ref{fig:rho224}. In contrast to the distinct pattern observed in the initial-state estimator, the final-state results do not show a statistically significant separation among the selected deformation configurations. This limitation is partly due to the high statistical demand of four-particle correlators, especially in a smaller heavy-ion system where the final state particle multiplicity is limited. Further investigation in larger systems and with higher event statistics is therefore required to establish the practical sensitivity of $\rho_{224}$ to the combined influence of quadrupole, triaxial, and hexadecapole deformations.

\begin{figure}[t]
  \centering
	\includegraphics[width=0.45\textwidth]{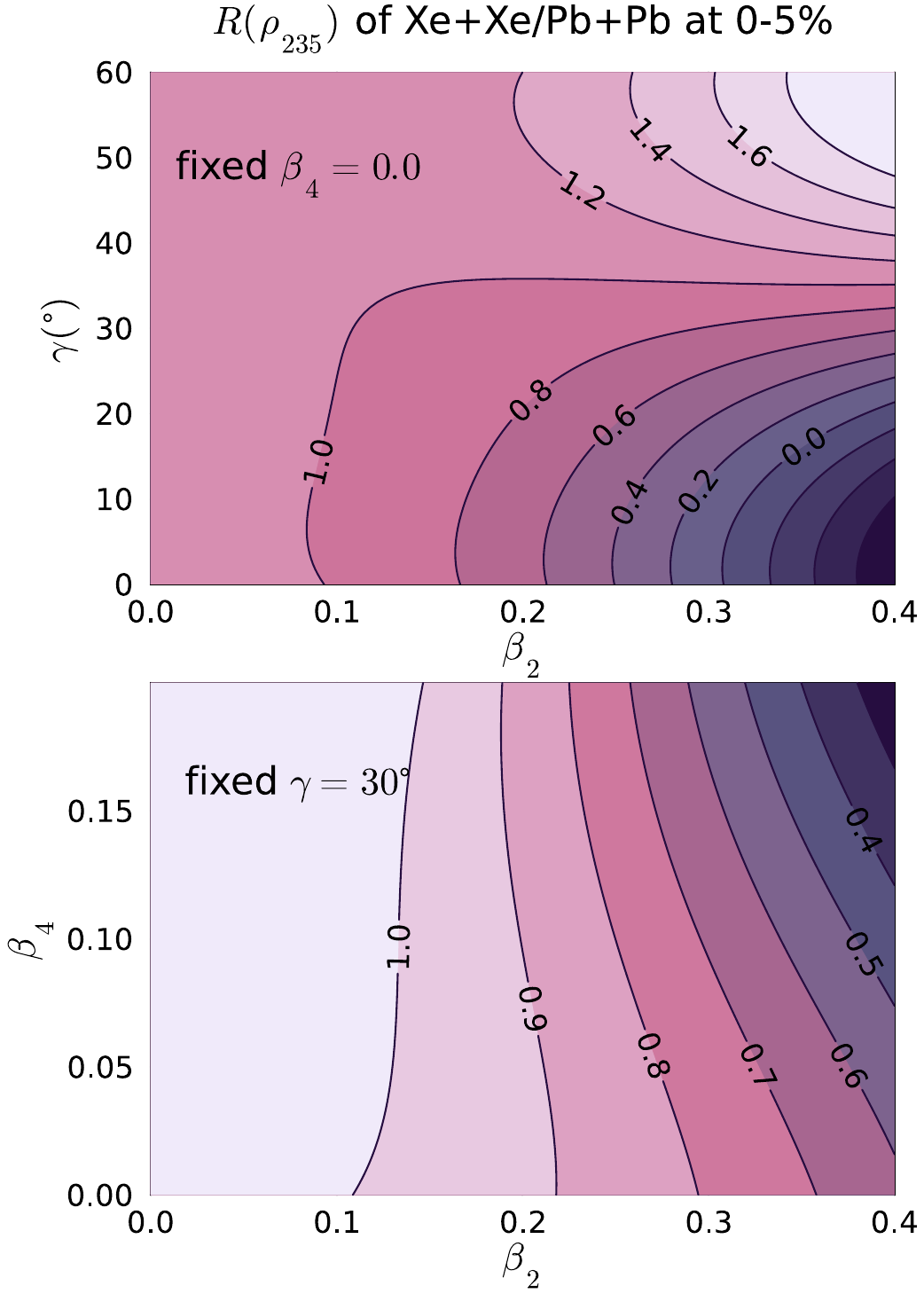}
	\caption{Contour plots of the initial-state estimator for the ratio $R(\rho_{235})$ in 0--5\% central $^{129}$Xe+$^{129}$Xe at $\sqrt{s_{\mathrm{NN}}}=5.44~\mathrm{TeV}$ over $^{208}$Pb+$^{208}$Pb collisions at $\sqrt{s_{\mathrm{NN}}}=5.02~\mathrm{TeV}$. The upper and lower panels show the slices on $(\beta_2, \gamma)$ plane at $\beta_4 = 0.0$ and $(\beta_2, \beta_4)$ plane at $\gamma = 30^\circ$, respectively.}
	\label{fig:rho235_contourf}
\end{figure}

We also examine another four-particle observable, $\rho_{235}$, defined in Eq.~(\ref{eq:rho_nmnm}). Fig.~\ref{fig:rho235_contourf} presents the initial-state contour slices of $R(\rho_{235})$. In the upper $(\beta_2, \gamma)$ panel, $R(\rho_{235})$ shows a pronounced dependence on both $\beta_2$ and $\gamma$. By contrast, the lower $(\beta_2, \beta_4)$ panel indicates a more limited sensitivity to $\beta_4$. This observation is quantitatively supported by the distance-correlation analysis, $\mathrm{dCor}(\rho_{235},\beta_2)=0.568$, $\mathrm{dCor}(\rho_{235},\gamma)=0.656$, and $\mathrm{dCor}(\rho_{235},\beta_4)=0.077$. Similar to the case of $\rho_{224}$, the limited particle multiplicity and the associated statistical fluctuations make it difficult to conclusively resolve the nuclear deformation effects in the final-state $\rho_{235}$.

\subsection{\texorpdfstring{$v_n^2-v_m^2$}{vn squared--vm squared} correlation}

\begin{figure}[t]
	\centering
	\includegraphics[width=0.45\textwidth]{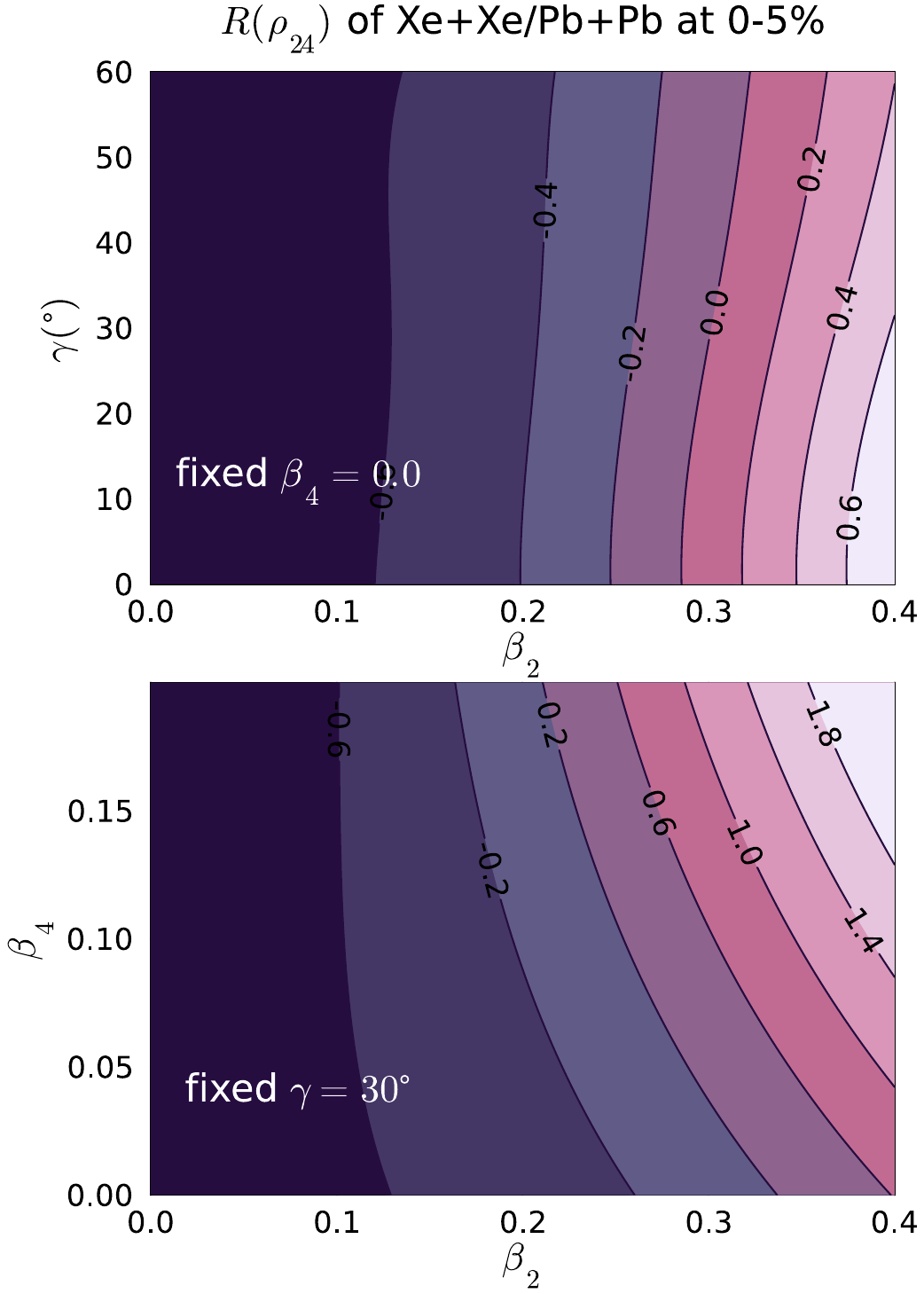}
	\caption{Contour plots of the initial-state estimator for the ratio $R(\rho_{24})$ in 0--5\% central $^{129}$Xe+$^{129}$Xe at $\sqrt{s_{\mathrm{NN}}}=5.44~\mathrm{TeV}$ over $^{208}$Pb+$^{208}$Pb collisions at $\sqrt{s_{\mathrm{NN}}}=5.02~\mathrm{TeV}$. The upper and lower panels show the slices on $(\beta_2, \gamma)$ plane at $\beta_4 = 0.0$ and $(\beta_2, \beta_4)$ plane at $\gamma = 30^\circ$, respectively.}
	\label{fig:rho24_contourf}
\end{figure}

\begin{figure}[t]
	\centering
	\includegraphics[width=0.45\textwidth]{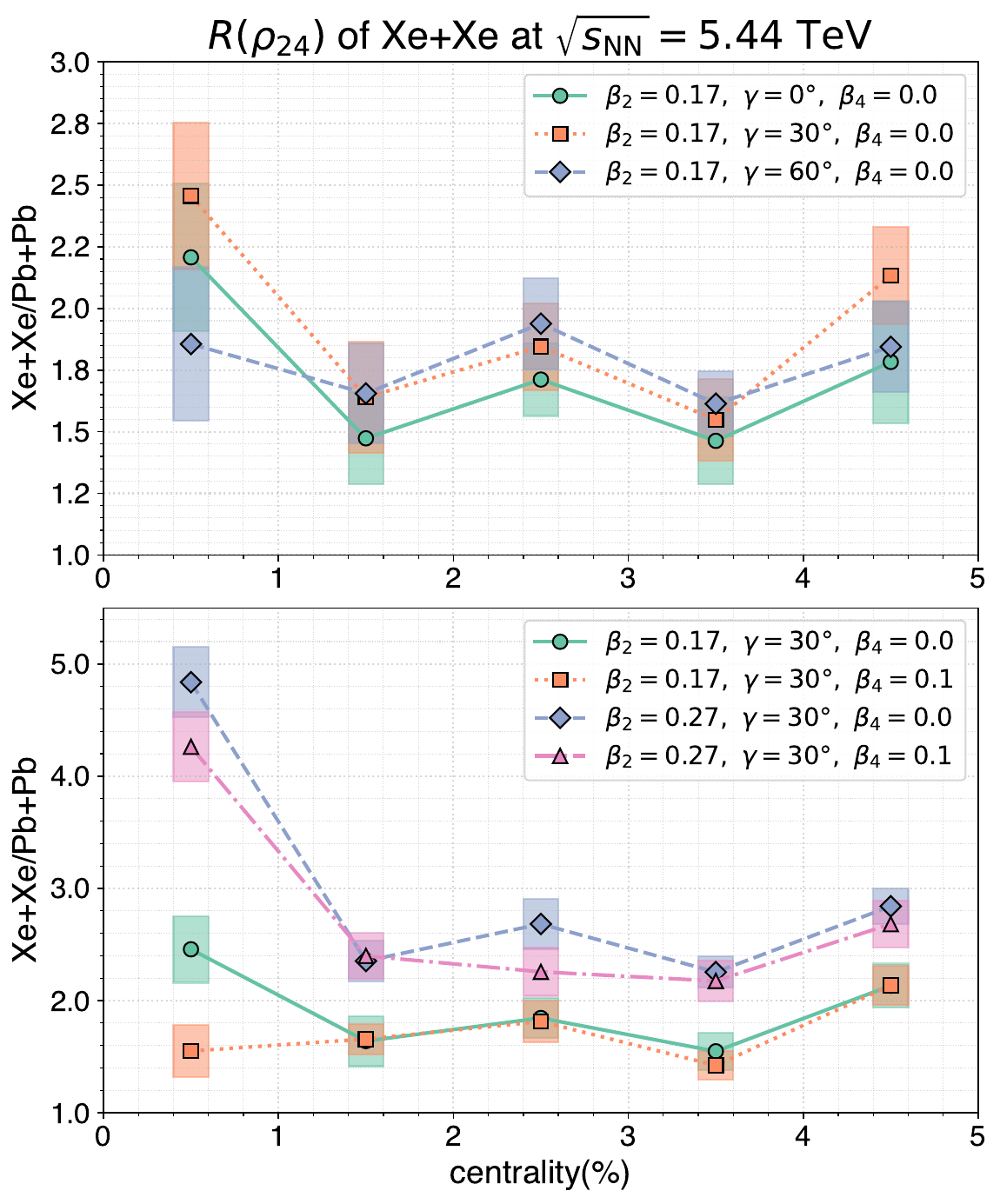}
	\caption{Centrality dependence of $R(\rho_{24})$ calculated by iEBE-VISHNU model. The upper panel shows the effect of varying $\gamma$ at fixed $\beta_2 = 0.17$ and $\beta_4 = 0.0$, while the lower panel compares configurations with different $\beta_2$ and $\beta_4$ values at fixed $\gamma = 30^{\circ}$.}
	\label{fig:rho24}
\end{figure}

The four-particle correlation between $v_n^2$ and $v_m^2$, denoted as $\rho_{nm}$, is investigated in this section. Fig.~\ref{fig:rho24_contourf} shows the 2D contour slices of the initial-state estimator for $R(\rho_{24})$. The observable is dominated by the quadrupole deformation, as reflected by the strong increase of $R(\rho_{24})$ with $\beta_2$ in both panels. The distance-correlation values for global sensitivity analysis, $\mathrm{dCor}(\rho_{24},\beta_2)=0.911$, $\mathrm{dCor}(\rho_{24},\gamma)=0.194$, and $\mathrm{dCor}(\rho_{24},\beta_4)=0.219$, confirm this dominant $\beta_2$ dependence. Nevertheless, when $\beta_2$ becomes large, the contour lines are visibly tilted along both the $\gamma$ and $\beta_4$ directions. In particular, at $\beta_2\simeq0.35$ and $\gamma=30^\circ$, increasing $\beta_4$ from 0 to 0.2 raises $R(\rho_{24})$ from roughly 0.4 to 1.4, indicating that the $\beta_4$ effect becomes appreciable only in the presence of a large quadrupole deformation.

Fig.~\ref{fig:rho24} presents the iEBE-VISHNU hybrid simulation results for $R(\rho_{24})$. The observable suffers from considerable statistical fluctuations, as expected for a four-particle correlator in the medium-size $^{129}$Xe system. In the upper panel, the results for different $\gamma$ with fixed $\beta_2=0.17$ are indistinguishable within statistical uncertainties, which is consistent with the weak $\gamma$ dependence predicted by the initial-state estimator in $\beta_2<0.2$ region. In the lower panel, despite the statistical noise, the curves for $\beta_2=0.27$ are distinctly separated from those for $\beta_2=0.17$, proving the dominant role of $\beta_2$. By contrast, the $\beta_4$ splitting predicted at large $\beta_2$ in the initial state is not resolved within the current statistical uncertainties.

\subsection{SHapley Additive exPlanations}

\begin{figure}[t]
	\centering
	\includegraphics[width=0.45\textwidth]{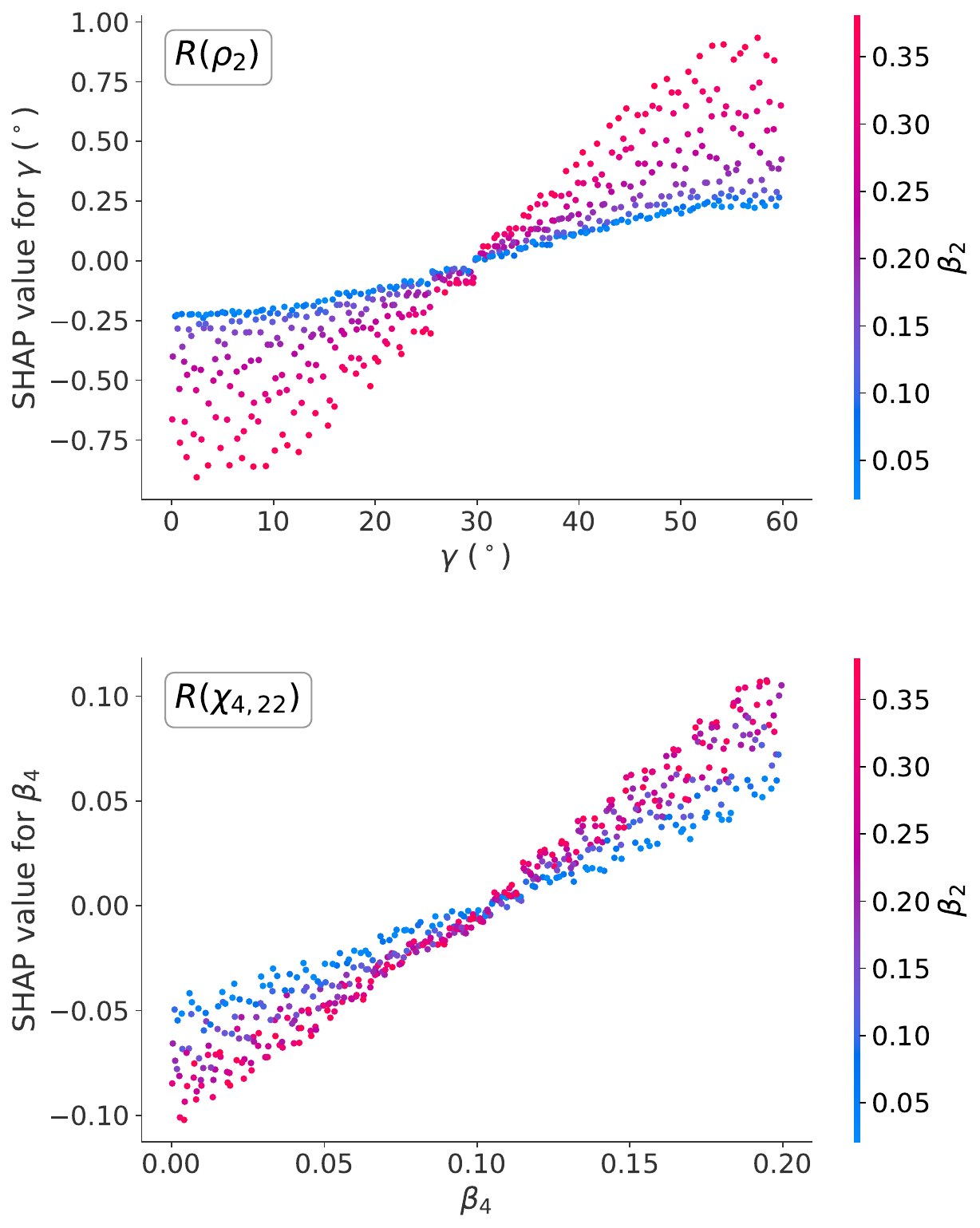}
	\caption{SHAP dependence plots of the initial-state observables $R(\rho_2)$ as a function of $\gamma$ (upper), and $R(\chi_{4,22})$ as a function of $\beta_4$ (lower). In both panels, the scatter points are color-coded by $\beta_2$ to visually illustrate how the SHAP values associated with $\gamma$ and $\beta_4$ vary with $\beta_2$.}
	\label{fig:SHAP}
\end{figure}

While two-dimensional contour slices are useful for the present three-parameter study, they become inadequate in higher-dimensional deformation spaces, where the multidimensional dependence of observables on these parameters requires a more systematic characterization. This limitation motivates the introduction of the SHapley Additive exPlanations (SHAP) method as a complementary tool \cite{SHAP1,SHAP2}. By decomposing the model prediction into additive marginal contributions from individual input parameters, SHAP provides a quantitative and scalable framework to assess parameter sensitivities and their interplay in high-dimensional spaces \cite{SHAP2}. For a prediction model $f(\mathbf{x})$ with parameter configuration $\mathbf{x}=(x_1,\dots,x_M)$, the SHAP representation can be written as \cite{SHAP1}
\begin{equation}
    f(\mathbf{x}) = \phi_0 + \sum_{i=1}^{M}\phi_i(\mathbf{x}),
\end{equation}
where $\phi_0$ is the average prediction over the sampled parameter space. The SHAP value $\phi_i(\mathbf{x})$ quantifies the contribution of the $i$-th parameter, with a positive (negative) $\phi_i$ indicating an enhancement (suppression) of the observable relative to $\phi_0$. We emphasize that SHAP is not used here as a Bayesian inference framework to constrain physical input parameters from data. Instead, it serves as a diagnostic sensitivity tool, quantifying how variations in these inputs contribute to the calculated observable response.

To illustrate how SHAP characterizes the dependence of observables on multiple parameters, we revisit two examples discussed before: the dependence of $\rho_2^{(\mathrm{init})}$ on the triaxiality $\gamma$ and that of $\chi_{4,22}^{(\mathrm{init})}$ on the hexadecapole deformation $\beta_4$. Their SHAP dependence plots are shown in Fig.~\ref{fig:SHAP}, with each point corresponding to one sampled nuclear shape configuration in the parameter space. The horizontal axis denotes the value of the primary parameter, namely $\gamma$ for $\rho_2^{(\mathrm{init})}$ and $\beta_4$ for $\chi_{4,22}^{(\mathrm{init})}$, while the vertical axis shows the associated SHAP value. The points are color-coded by $\beta_2$, allowing the combined influence of $\beta_2$ to be tracked.

At a fixed primary parameter, the vertical spread of the points reflects residual dependence on the remaining parameters. The color ordering therefore provides a qualitative indication of how much of this dispersion is associated with $\beta_2$. In the upper panel, the strong color ordering reveals a strong interaction effect between $\gamma$ and $\beta_2$ on $\rho_2^{(\mathrm{init})}$. In the lower panel, the narrower color separation indicates a weaker modulation of the $\beta_4$ contribution to $\chi_{4,22}^{(\mathrm{init})}$ by $\beta_2$, consistent with the contour-plot observation that $\chi_{4,22}$ is predominantly governed by $\beta_4$.

\section{Summary}
\label{sec:summary}

In this work, we have performed a joint analysis of nuclear deformation effects on multiparticle observables in relativistic heavy-ion collisions within a three-dimensional parameter space spanned by $(\beta_2,\gamma,\beta_4)$. Using an initial-state framework based on a modified Monte Carlo Glauber model, complemented by hybrid simulations with iEBE-VISHNU, we examined the response of various observables to combined variations of deformation parameters, taking $^{129}$Xe+$^{129}$Xe collisions at $\sqrt{s_{\mathrm{NN}}}=5.44~\mathrm{TeV}$ as a representative example.

At the initial-state level, our analysis shows that $\rho_2$ is strongly sensitive to $\beta_2$ and $\gamma$, but remains largely insensitive to $\beta_4$. Especially, the $\gamma$ dependence becomes more visible at larger $\beta_2$. For the nonlinear response coefficient $\chi_{4,22}$, we find that the initial-state sensitivity is dominated by $\beta_4$, while the dependence on $\beta_2$ and $\gamma$ remains comparatively weak. We also examined higher-order four-particle correlators, including $\rho_{224}$, $\rho_{235}$, and $\rho_{24}$, whose initial-state estimators exhibit distinct multidimensional sensitivity patterns. In particular, $\rho_{224}$ is strongly sensitive to $\beta_2$, while also showing a combined dependence on $\gamma$ and $\beta_4$ in the presence of a finite $\beta_2$ background. These results demonstrate that the geometric response of multiparticle observables cannot, in general, be inferred from isolated one-parameter variations alone.

At the final-state level, the simulation results indicate that the sensitivity of $\rho_2$ to quadrupole deformation and triaxiality is largely preserved in $^{129}$Xe+$^{129}$Xe collisions, supporting its robustness as a practical probe of these deformation features. By contrast, the strong initial-state sensitivity of $\chi_{4,22}$ to $\beta_4$ is not maintained after the full dynamical evolution, where its variation becomes difficult to distinguish from background fluctuations within the present statistical precision. Similarly, although the higher-order correlators display nontrivial deformation sensitivity at the initial-state level, their corresponding final-state signals cannot be resolved conclusively in the present hybrid calculations because of limited event statistics. These comparisons show that initial-state sensitivity alone is not sufficient to establish the practical probing capability of a final-state observable. At the same time, the initial-state analysis provides useful guidance for identifying observables whose geometric sensitivity may survive the subsequent dynamical evolution. Extending such studies to larger collision systems would be particularly useful, where higher multiplicities and stronger collective flow may help reduce statistical fluctuations and make deformation-driven signals easier to resolve.

Overall, our results indicate that the deformation sensitivity of multiparticle observables is governed not only by individual deformation parameters, but also by their combined variations within the multidimensional deformation space. A joint analysis is therefore necessary for assessing whether an observable provides a robust deformation constraint or only responds within a restricted region of parameter space.

\begin{acknowledgments}
We acknowledge Prof. Haojie Xu for useful discussions and comments. We also thank Dr. Shujun Zhao and Dr. Lu-Meng Liu for their helpful guidance on the technical details of the hydrodynamic simulations. This work was supported by the National Natural Science Foundation of China (NSFC) under Grant Nos. 12422508, 12375124, and the Science and Technology Commission of Shanghai Municipality under Grant No. 23JC1402700. Y.S. thanks the sponsorship from Yangyang Development Fund.
\end{acknowledgments}

\bibliographystyle{apsrev4-2}
\bibliography{references}

\end{document}